# Single-Shot Multispectral Encoding: Advancing Optical Lithography for Encryption and Spectroscopy

*Hyewon Shim[1,2], Geonwoong Park[2], Hyunsuk Yun[3], Sunmin Ryu[3], Yong-Young Noh[2]\*, and Cheol-Joo Kim[1,2]\**

[1]Center for Van der Waals Quantum Solids, Institute for Basic Science (IBS), Pohang 37673, Republic of Korea.

[2]Department of Chemical Engineering, Pohang University of Science and Technology (POSTECH), Pohang 37673, Republic of Korea.

[3]Department of Chemistry, POSTECH, Pohang 37673, Republic of Korea.

[\*]Corresponding Author: yynoh@postech.ac.kr, kimcj@postech.ac.kr

ABSTRACT

Most modern optical display and sensing devices utilize a limited number of spectral units within the visible range, based on human color perception. In contrast, the rapid advancement of machine-based pattern recognition and spectral analysis could facilitate the use of multispectral functional units, yet the challenge of creating complex, high-definition, and reproducible patterns with an increasing number of spectral units limits their widespread application. Here, we report a technique for optical lithography that employs a single-shot exposure to reproduce perovskite films with spatially controlled optical band gaps through light-induced compositional modulations. Luminescent patterns are designed to program correlations between spatial and spectral information, covering the entire visible spectral range. Using this platform, we demonstrate multispectral encoding patterns for encryption and multivariate optical converters for dispersive optics-free spectroscopy with high spectral resolution. The fabrication process is conducted at room temperature and can be extended to other material and device platforms.

MAIN TEXT

Optical sensors and displays play key roles in capturing and transporting optical information in modern information technology. As the number of units with varying spectral responses increases in devices, so too does the potential for multiplexed optical information within a broad bandwidth. However, the majority of devices are designed based on the trichromatic red-green-blue (RGB) color scheme, optimized for human color perceptions[1]. Recently, with the prevalence of machine-based analyses of optical patterns in both spatial and spectral domains, there is a significant demand for a multispectral optical platform[2,3]. This platform functions in two main roles: (*I*) transforming spectral information into spatially programmed patterns, and (*II*) translating geometric information into specific emissive spectra with programmed correlations between the spatial coordinates of the optical element and the optical spectra after light-matter interactions, as depicted in Fig. 1A. These capabilities enable the development of advanced optical and optoelectronic device applications, including multiplexing in optical communications[4,5], broadband displays exceeding the red-green-blue color model[6,7], and compact spectral analysis tools[8-10].

To manipulate the light spectrum across a broad band region through light-matter interactions, semiconductor alloys provide a compact, solid-state platform with continuously variable energy band gaps as their compositions change[11] (Fig. 1A). These alloys not only exhibit a broad spectrum range but can also be integrated within the same material platform to realize multispectral systems. Methods for both step-by-step[12,13] and simultaneous integrations[14,15] of alloys have been developed. However, spatially controlled integration becomes increasingly challenging as the number of spectral units required to extend multispectral bandwidth increases. Consequently, many promising ideas for multispectral functional elements remain at the proof-of-concept stage due to reproducibility challenges in fabricating identical structural copies through multiple integration processes. This often necessitates a fine calibration step after each fabrication to achieve the required optical properties on demand.

Here, we present a single-shot optical lithography technique for creating multispectral patterns in semiconductor alloys (Fig. 1B). Uniform semiconducting $CsPbBr_3$ perovskite films were deposited on substrates via solution processes at room temperature[16] (see fig. S2 for the film's characterization data). Single-shot optical lithography was performed on the film by illuminating it with the light of approximately 405 nm through a photomask in a controlled environment with vapor-phase precursors of dihalomethane ($CH_2X_2$, X = Cl, Br, I) in a custom-built chamber (fig. S1). A pre-designed metal mask with spatially varying thicknesses was used to modulate local light transmission from 0% to nearly 100% (fig. S3A), forming alloy patterns directly through controlled anion exchanges facilitated by the photo-induced decomposition of $CH_2X_2$[17,18], without the use of photoresist. This process allows numerous different alloy compositions to be simultaneously achieved in a single step at the desired locations, resulting in multispectral patterns with high spectral resolution and reproducibility.

Utilizing the designed optical density-modulated photomask (Fig. 1C), we have fabricated a multispectral fluorescent pattern that replicates a drawing of a cat with fishes on $CsPbBr_{3-x}I_x$ films. This was achieved where photo-induced decomposition of $CH_2I_2$ occurred in $CsPbBr_3$ films. The photoluminescence (PL) peak wavelength, $\lambda_{PL}$ mapping data (Fig. 1D), reveals the targeted features, thereby demonstrating the conversion of the optical density pattern into a spectral pattern. The multiple converted regions display programmed $\lambda_{PL}$, spanning from 523 nm to 606 nm, corresponding to the original $CsPbBr_3$ and $CsPbBr_{1.4}I_{1.6}$, respectively. Local PL spectra (fig. S3B-C), taken from various regions, exhibit singular PL peaks, indicating that homogeneous $CsPbBr_{3-x}I_x$ alloys without phase segregation have formed across the entire spectrum range, maintaining the direct bandgap nature[19]. Notably, no halogen exchange occurred without incident light, as evidenced by the identical $\lambda_{PL}$ after the process compared to the original CsPbBr$_3$ (fig. S2D). We also confirmed that the pattern could be replicated by repeating the process with the same multispectral mask (fig. S3D). While several reports[20,21] have discussed local composition modulations of perovskite alloys through post-synthetic anion exchanges, the number of possible spectral units was often limited due to integration issues, precluding the fabrication of large-scale multispectral patterns with high definition.

The material design platform is versatile, suitable for designing multi-band components that fulfill the two key functions described in Fig. 1A. To demonstrate function (*I*), we exhibit pattern encryptions, where $\lambda_{PL}$-dependent spatial patterns are encoded with precise spectral modulations. In Fig. 2A-C, the $\lambda_{PL}$ from local regions of the CsPbBr$_3$ film is slightly altered by controlling the incorporation of chloride elements with different gaseous precursors (see Methods), using the same lithography process with an optical density-graded photomask. Each programmed area displays a different PL spectrum with subtly tuned $\lambda_{PL}$ within 10 nm intervals (Fig. 2C). Selecting a certain bandwidth of PL (indicated by vertical shadowed pink lines in Fig. 2C) with bandpass filters, the relative PL intensity, $I_{PL}$, from different regions changes, creating bandwidth-dependent fluorescent intensity patterns, as shown in Fig. 2D. Employing this mechanism, we patterned multiplexed fluorescent patterns of numeric codes in a CsPbBr$_{3-x}$Cl$_x$ perovskite film (Fig. 2E). Given that the spectral modulation is precisely conducted within a narrow bandwidth, it is difficult for conventional sensor and display units with Bayer filters to capture and reproduce the fluorescent patterns (Fig. 2E, bottom left). However, when bandpass filters with narrow 1 nm bandwidths are used at different center wavelengths, wavelength-dependent numeric codes become visible as "1515" at 518 nm, "7976" at 520 nm, and "3808" at 525 nm. This demonstrates that multiple $\lambda_{PL}$-dependent patterns can be employed to thwart decryption attempts by incorporating numerous false patterns.

As the spectra are determined by the intrinsic optical band gaps of semiconductors, the $\lambda_{PL}$ can be tuned within a desired wavelength range to form patterns of arbitrary shapes and sizes in the macroscopic regime. To demonstrate the versatility of our process, we patterned a QR code linking to the web page of our laboratory by slightly modulating the PL spectra in $CsPbI_xBr_{3-x}$ perovskite films. The binary pattern is constructed using two spectral units with slightly different $\lambda_{PL}$ of 545 nm and 557 nm, respectively (Fig. 3A), with a central wavelength peak of approximately 550 nm, emitting yellow fluorescent light (Fig. 3B). Hue purity discrimination by color perception is known to be weakest for yellow light against a white background for the majority of people, as the most sensitive M- and L-cone cells are similarly activated by both yellow and white light[22]. Therefore, it is challenging to recognize the pattern with the naked eye or conventional sensors with Bayer filters, and the designed pattern was revealed in high contrast only when the correct spectral filter was applied (Fig. 3C). We note that the PL spectra are maintained regardless of the viewing angle or light source, generating PL-dependent fluorescent patterns that are recognizable by image sensors (Fig. 3D), unlike the case with spectral filters used to create structural colors[23,24]. These measurements confirm that encryption can be achieved on surfaces with curvature, such as flexible substrates, for a broad range of applications[25,26]. The pattern can be maintained for several months, demonstrating the long retention time, despite of intermixing between two phases in the binary pattern (fig. S5). Also, the pattern can be erased for writing a new pattern in the same film (fig. S6).

Our encryption approach offers several significant advantages over previous methods[27-31]. Firstly, cryptographic security is enhanced because the optical information can only be decoded by selecting the appropriate filter bandwidth. Material systems for optical encryption are generally fabricated by patterning a specific chemical substance, designed to provide optical contrast with the surroundings and reveal hidden patterns when subjected to targeted external stimuli, such as temperature changes[27,28], moisture[29], and chemical reactions with solvents[30,32]. However, the responsiveness of the patterned chemicals can sometimes exhibit poor selectivity to these stimuli. For instance, the optical properties of molecules are frequently influenced by interactions with various molecules, which increases the risk of information leakage through unauthorized decryption attempts. In contrast, our method encodes patterns through subtle changes in chemical composition, making them challenging to decipher without precise bandwidth information.

In addition, decoding the fluorescent pattern is a non-destructive process, unlike invasive methods where the material platform undergoes chemical or structural change to become optically active[28-30]. Consequently, our films are suitable for the development of optoelectronic devices capable of dynamically generating various encryption patterns on a single platform. For example, an array of green-light emitting diodes could be created, each with slight spectral shifts that are only detectable using the correct bandpass filters. Given the heightened risk of image information exposure due to the prevalent use of electromagnetic wave sensors in contemporary society[32], this platform offers a promising avenue for enhancing security in optoelectronic devices beyond the traditional red/green/blue color scheme.

We now demonstrate Function (*II*) of the programmable multispectral emissive elements, which tune emission spectra by manipulating the coordinates defining the PL emission area (Fig. 4A). Whereas monochromators and interferometers achieve the same functionality by utilizing moving optical elements, such as gratings and mirrors, to select the desired band from a broadband light source, our method generates multivariant PL spectra through the down-conversion of single-wavelength excitation light at each position. Consequently, this multivariant optical film can be paired with a continuous-wave laser, rather than bulky lamps, to create a wavelength-tunable light source for optical spectroscopy. Given that the excitation laser can be focused to diffraction limits, the emitted light from a small size sample can cover a broad spectral range. This is achieved by raster-scanning the confocal excitation over a compact film, wherein different semiconductor alloys are densely integrated, facilitating the realization of a miniaturized spectrometer (Fig. 4B-C).

We fabricated continuous composition-graded $CsPbCl_{3-x}Br_x$ and $CsPbI_xBr_{3-x}$ alloy films with $x$ values, ranging from 0 to nearly 3 along a fixed direction. A continuously variable optical density filter served as the photomask for multispectral lithography, enabling the translation of optical density information into spectral information (Fig. 4A). The local PL spectra, obtained by a confocal microscope, exhibit single peaks with position-dependent $\lambda_{PL}$, spanning the entire visible spectral range (Fig. 4D). This indicates the absence of abrupt changes in optical band gaps within the scanned area. In this process, all the spectral units are integrated within a narrowly defined region, only 175 μm wide, a feat challenging to accomplish using conventional step-by-step integration methods[8,12,13,20].

This system serves as a dispersive optics-free spectrometer for optical characterizations. After characterizing the PL properties of the multivariant emissive film to obtain the three-dimensional plot of $I_{PL}$ (Fig. 4D), which can be represented by an m-by-n transfer matrix, A, with rows and columns corresponding to spatial and spectral coordinates respectively (Fig. 4C), the position-dependent $I_{PL}(a)$ is measured through a sample of interest using a single-cell photodetector without a spectrometer. If $T_{sample}(\lambda)$ represents an n-by-1 matrix for the optical transmission of PL signals at each wavelength through the sample, then it can be deduced by optimization fitting to minimize A · $T_{sample}(\lambda) - I_{sample}(a)$, where $I_{sample}(a)$ represents an m-by-1 matrix for the position-dependent $I_{PL}$ (refer to the Supporting Information for a detailed process). We used two narrow bandpass filters as reference samples to validate our approach and confirmed that the deduced $T_{sample}(\lambda)$ (red color spectra in Fig. 4E-F) closely matches the transmission spectra measured with a conventional monochromator (black spectra). Notably, as m, the number of spectra with different $\lambda_{PL}$ at various positions can exceed n, the number of spectral values, which is limited by the spectral resolution of the spectrometer used to acquire A, one can maintain high spectral resolution to detect the narrow bandwidth of the references through the optimization process.

Our approach to design multispectral patterns can be versatile to apply to different semiconductor alloy systems with right selections of the host film and gaseous precursor. The exchange process primarily consists of two steps: (*i*) photocatalytic decomposition of the vapor precursor on the film surface and (*ii*) subsequent diffusion of decomposed elements into the bulk crystal (Fig. 5A)[33]. Step (*i*) should be the rate-determining step to form an intermixed ternary alloy throughout the film and tune the $\lambda_{PL}$ by adjusting the photon flux during the photoconversion of constituent elements. The rate of each step can vary dramatically based on the strength of the photocatalytic effect of the host film in decomposing adjacent precursors and the diffusion coefficient of the incorporated elements within the film, determining the overall process.

The PL spectra modulation in $CsPbCl_{3-x}Br_x$ film strongly depends on the type of host film, even under identical illumination conditions. When Br substitutes for Cl in $CsPbCl_3$ using a $CH_2Br_2$ precursor (Fig. 5B), single peaks with continuously varying $\lambda_{PL}$ from 420 nm to 523 nm emerge. Conversely, when Cl replaced Br in $CsPbBr_3$ using a $CH_2Cl_2$ precursor, the PL peak for $CsPbBr_3$ at 520 nm gradually diminishes, while the peak for $CsPbCl_3$ at 420 nm appears (Fig. 5C), indicating the formation of segregated phases of $CsPbBr_3$ and $CsPbCl_3$. The differing behaviors in these cases can be attributed to the varying halogen diffusivities in the host films, influenced by atomic vacancies[34-36]. $CsPbBr_3$ has been reported to possess a significantly lower equilibrium vacancy concentration than $CsPbCl_3$, with vacancy formation energies for Schottky defect pairs estimated at 0.35 eV and 0.42 eV for $CsPbBr_3$ and $CsPbCl_3$, respectively[34]. As a result, the interdiffusion rate of Cl into $CsPbBr_3$ crystallites from the surface is significantly reduced, leading to an abrupt junction (see schematics in Fig. 5C). Since intermixed perovskite alloys are more thermodynamically stable than segregated phases[37,38], Cl will eventually diffuse throughout the $CsPbBr_3$ crystallites. Nevertheless, if the decomposition rate of Cl-containing precursors on the surface exceeds the diffusion rate, the resulting composition profile will be diffusion-limited with an abrupt junction.

For the case of Fig. 5B, the relative ratio of substituted Br among all the halogens, $X_{Br}$, inferred from $\lambda_{PL}$ is plotted as a function of the photo-excited carrier density, $N_{ex}$ at the surface of the host film (Fig. 5D). Assuming the total process limited by step ($i$), we fitted the data to an asymptotic exponential function representing saturated exchange behavior as $1 - e^{-\alpha \frac{N_{ex}}{d_h}}$, where $d_h$ is the areal density of the halogen elements at the surface, and $\alpha$ indicates the photocatalytic efficiency for halogen exchanges (see Supporting Information for details). The deduced α value is close to 2 in the low $N_{ex}$ regime (blue fitted curve), indicating the efficient photoconversion of the dihalide precursor of $CH_2Br_2$. As the conversion progresses, the halogen conversion rate by decomposing the $CH_2Br_2$ precursor slows, with the reduced α value of 0.53 (red curve). This slowdown is presumably due to a change in the relative electronic energy levels of the Br-rich $CsPbCl_{3-x}Br_x$ film in the photo-excited states, and the precursor molecules, which govern the charge transfer process essential for photocatalytic conversion[18,39]. Nevertheless, the entire process is constrained by the slower step ($i$), enabling precise modulation of $\lambda_{PL}$ by controlling the excited UV photon flux to achieve full conversion with $X_{Br}$ approaching 1.

The α value is highly sensitive to the type of precursor as well, affecting the kinetics of the overall process. When iodine was substituted for bromine in $CsPbBr_3$ using an $CH_2I_2$ precursor, modulation of $\lambda_{PL}$ from a single peak was observed, indicating the formation of homogeneous $CsPbBr_{3-x}I_x$ alloys (Fig. 4D, fig. S3C). Unlike chlorine, which separates phases within the same host film, iodides form a mixed phase because iodine has a faster diffusion rate than chlorine within the $CsPbBr_3$ structure. This is due to their relative bonding strength to lead, with chlorine forming a stronger bond than iodine[40]. In the $N_{ex}$-dependent $X_I$ data (Fig. 5E), the deduced α of $8.4 \times 10^{-3}$ is more than two orders of magnitude lower than that in Fig. 5D. This finding aligns with the higher dissociation energy of C-I bonds in the $CH_2I_2$ precursor than C-Br bonds in the $CH_2Br_2$ precursor[41]. Thus, the slower step (*i*) could still be the rate-determining step despite the expectedly low Iodine-diffusion rate[41] in the $CsPbBr_3$ film of $2.08 \times 10^{-11}$ cm$^2$/s. We note that the diffusivity plays important roles to determine not only the kinetics of halide exchange process, but also the spatial resolution of our technique (fig. S7), and can be engineered by changing the grain structures of host films (refer to the Supporting Information for a detailed discussion).

Our approach to designing semiconductor alloy patterns can be extensively applied across a range of applications. Lead-based halide perovskites are known to be excellent hole and electron conductors[42,43], allowing for the creation of composition-modulated circuitry that forms various functional multispectral optoelectronic elements. These include wavelength-tunable solid-state light emitters with confined type-I heterostructures, and broadband photodiodes with band-graded junctions. Alloy formation and patterning occur simultaneously through a bottom-up approach, eliminating the need for additional integration steps; thus, the fabrication of complex integrated circuitry is made possible. Furthermore, by using different precursors, this method can be adapted to various semiconductor systems[44] for spatial control of their band gap and/or electrochemical potentials. This strategy is particularly advantageous for designing circuitry with nanomaterial systems[11,45], where conventional top-down patterning processes are not viable due to structural or chemical instability.

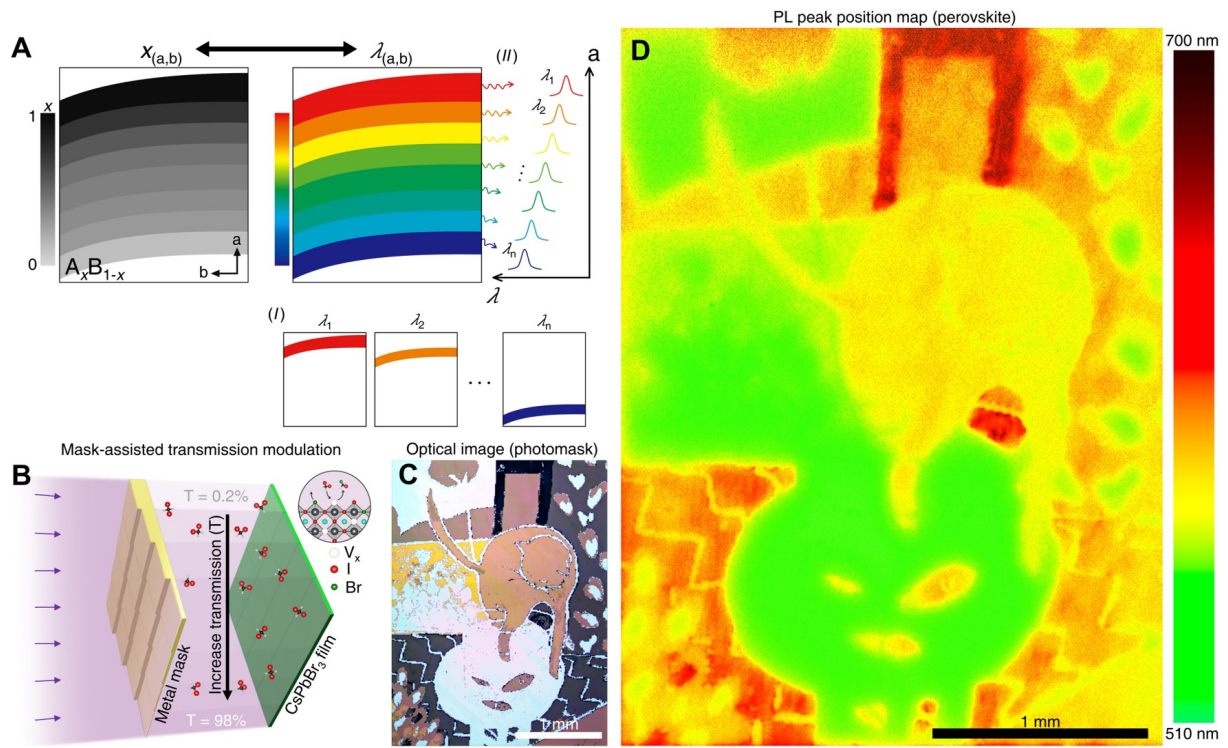

**Figure 1. Single-shot Optical Lithography for Multispectral Patterns.** (A) Schematics for a multispectral optical platform with programmed correlations between semiconductor alloy composition and optical spectra. (B) Illustration of single-shot optical lithography process for multispectral patterning with optical density-graded photomask. (C) Optical reflectance image of a photomask, replicating a drawing of a cat with fishes. (D) PL peak position map of $CsPbI_xBr_{3-x}$ perovskite film after multispectral patterning.

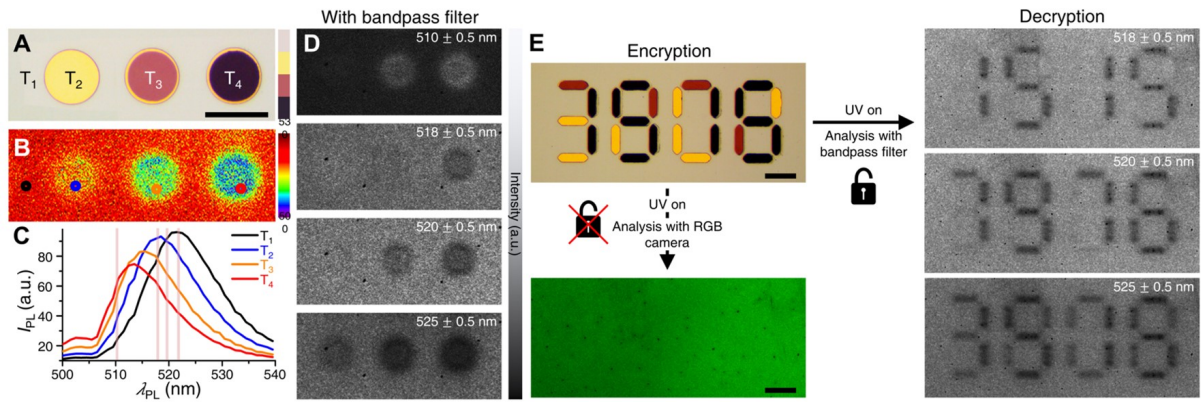

**Figure 2. Multispectral Optical Encryption.** (A) Optical reflectance image of a density-graded photomask. Scale bar: 50 μm. (B) PL peak position map of patterned $CsPbCl_xBr_{3-x}$ perovskite, using the photomask shown in (A). (C) Local PL spectra corresponding to the circled areas in (B). (D) Fluorescence images captured with bandpass filters, whose bandwidths are shown in (C) with shadowed pink vertical lines. (E) Optical encryption of numeric codes patterned in perovskite. The bottom-left image is the fluorescent image under UV illumination, taken with a conventional RGB sensor. The right images show wavelength-dependent PL images through bandpass filters with different bandwidths, indicated at the right top corners.

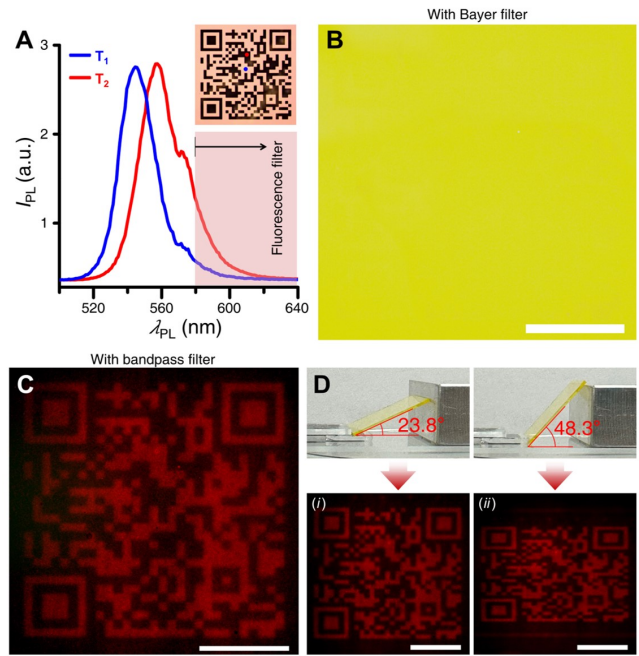

**Figure 3. Encrypted PL Patterns with Tunable Center Wavelengths.** (A) PL spectra from binary $CsPbI_xBr_{3-x}$ patterns of a QR code (inset). The shadowed area indicates the transmitted range of the longpass filter used in (C) and (D). The central wavelength peak, determined by the intrinsic optical band gap, is approximately 550 nm, emitting yellow fluorescent light. (B) Fluorescence image of the patterned perovskite, taken by an image sensor with a Bayer filter. (C) Fluorescence image of the same film through the longpass filter. (D) (top) Photographs of the film at different tilt angles, and (bottom) fluorescence images through the longpass filter. The same pattern is maintained regardless of the viewing angle. All scale bars: 500 μm.

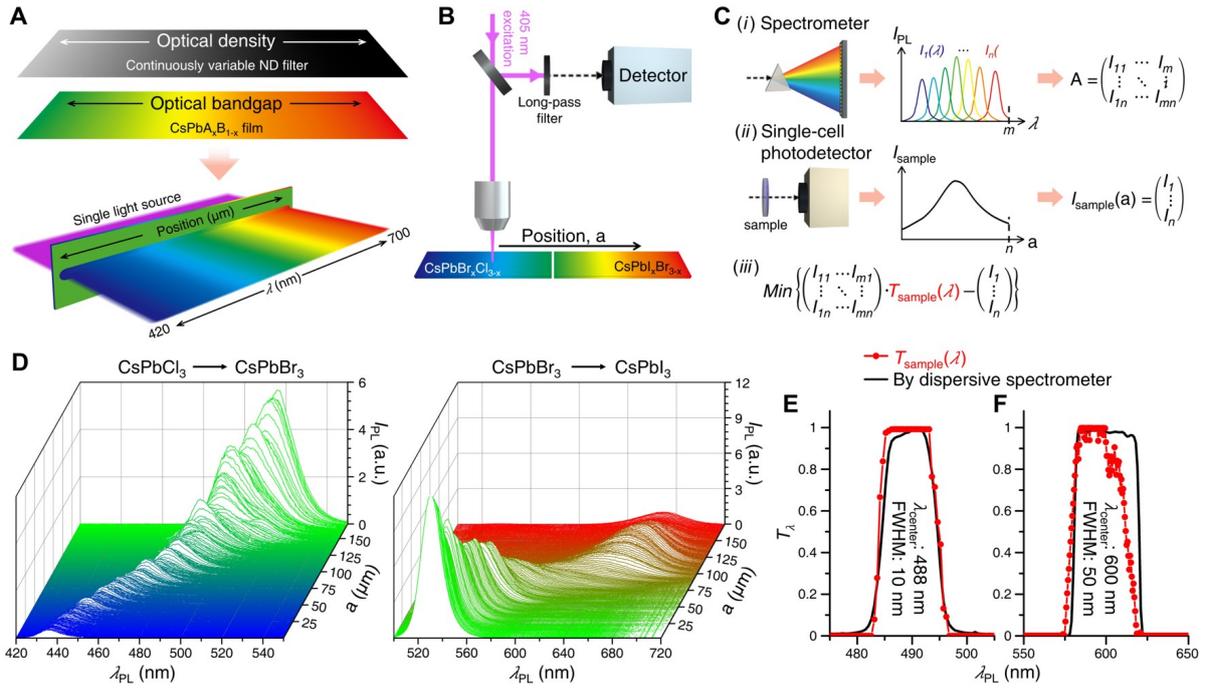

**Figure 4. Compressive Optical Spectroscopy with Multispectral Emission Films.** (A) Schematic of a multispectral emission film, fabricated by translating optical density information into spectral information. The emission spectra from a single light source are tuned by manipulating the spatial coordinates. (B) Schematic of a miniaturized spectrometer. (C) The operations of compressive spectroscopy. (D) Position-dependent local PL spectra of $CsPbBr_xCl_{3-x}$ and $CsPbI_xBr_{3-x}$. (E) and (F) is two bandpass filters with different bandwidths are used as the reference samples, whose transmittance data are shown in black. The deduced $T_{sample}(\lambda)$ by the compressive optical spectroscopy is shown in red.

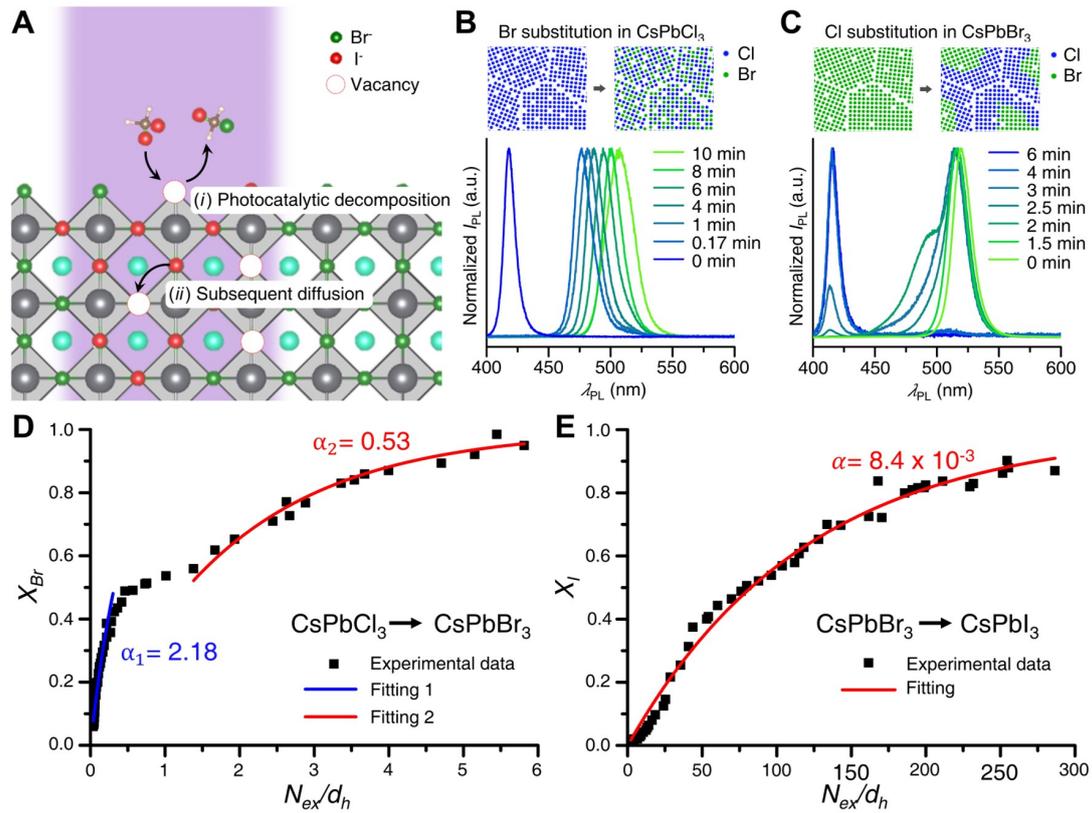

**Figure 5. Kinetics for Photocatalytic Conversion of Halides in Semiconductor Alloys.** (A) Schematic of the halide anion exchange process in perovskites. (B) and (C) is PL spectra with different reaction times for (D) Br replacement with Cl, starting from $CsPbCl_3$, and (C) Cl replacement with Br, starting from $CsPbBr_3$. (D) and (E) The composition ratio of the substituted halogen anion, $X_H$ (H = Br or I), as a function of photo-excited carrier density, $N_{ex}$, normalized by the areal density of halogen elements at the surface, $d_h$. Solid lines represent the fitting based on an asymptotic exponential function representing saturated exchange behavior, where α indicates the photocatalytic efficiency for halogen exchanges.

## ASSOCIATED CONTENT

**Supporting Information**

Preparation of host perovskite films, Fabrication of optical density-gradient photomasks, Single-shot optical lithography for multispectral patterns, Optical characterizations of multispectral patterns, Demonstrations of compressive optical spectroscopy with multivariant emissive films, Estimation of photocatalytic efficiency for halide exchange, α. (PDF)

## AUTHOR INFORMATION

**Corresponding Author:** yynoh@postech.ac.kr, kimcj@postech.ac.kr

**Author Contributions**

H.-W.S., Y.-Y.N. and C.-J.K. designed the experiments. G.-W.P. synthesized $CsPbBr_3$ samples. H.-W.S. built the optical lithography set-up and fabricated multispectral patterns. H.-W.S. and H.-S.Y. conducted PL measurements. H.-W.S. conducted optical data analyses for the demonstration of dispersive optics-free spectroscopy. H.-W.S. and C.-J.K. wrote the manuscript with inputs from all authors.

**Funding Sources**

This research was supported by the National R&D Program through the National Research Foundation of Korea (NRF) funded by the Ministry of Science and ICT (2020R1A4A1019455, 2020M3D1A1110548, 2023R1A2C2005427) and the Institute for Basic Science (IBS-R034-D1).

**Competing interests**

The authors declare no conflict of interest.

Supporting information

### Preparation of host perovskite films

A 0.35 M CsPbBr$_3$ precursor solution was prepared by mixing 74 mg of CsBr and 128 mg of PbBr$_2$ in dimethyl sulfoxide solvent, followed by heating at 60 °C with stirring overnight. The perovskite solution was then filtered using a polytetrafluoroethylene syringe filter with a pore size of 0.2 μm. The growth substrates were treated with ultraviolet (UV)/ozone for 30 min to make the surface hydrophilic. Subsequently, 35 μL of the precursor solution was spin-coated onto the substrates at 3000 rpm for 60 s. After coating, the perovskite films were annealed at 70 °C for 10 min. All precursor solutions and films were prepared in an N$_2$-filled glove box (O$_2$ and H$_2$O levels less than 10 ppm). To form CsPbI$_x$Br$_{3-x}$ alloy films, the as-grown CsPbBr$_3$ was used as the host film. For CsPbBr$_x$Cl$_{3-x}$ alloy films, the CsPbBr$_3$ film was first converted to CsPbCl$_3$ by the photo-induced halogen exchange process with CH$_2$Cl$_2$ precursors, then the CsPbCl$_3$ film was used as the host film.

### Fabrication of optical density-gradient photomasks

Photoresist (DNR) was spin-coated at 4000 rpm for 60 s on a fused silica substrate. The pre-baking process occurred on a hotplate at 90 °C for 90 s, followed by illumination with UV light for 17 s to pattern the desired figure. The patterned substrate was then post-baked on a hotplate at 100 °C for 60 s. The film was subsequently dipped in the developer for 35 s, rinsed with DI water, and dried with blowing air. After development, masking metals (Cr or Au) were thermally evaporated onto the patterned area with the targeted thickness depending on the desired optical transmittance (see fig. S3A) in a high vacuum chamber ($\sim10^{-6}$ Torr). Finally, the photoresist layer was lifted off using acetone. This entire process was repeated multiple times using a contact aligner to form patterns with units exhibiting different optical transmittance.

**Single-shot optical lithography for multispectral patterns**

Two quartz containers with 3 mL of $CH_2X_2$ (X: Cl, Br or I) liquids were loaded inside the optical chamber beside the $CsPbBr_3$ film at the center (fig. S1A). The density-graded optical mask was placed on top of the perovskite film. The chamber was sealed in a glove box of the $N_2$ atmosphere and then transferred to an optical stage. A laser beam of 405 nm passed through a beam expander to illuminate the entire area of the optical mask for the desired times.

**Optical characterizations of multispectral patterns**

Hyperspectral images of perovskite films following single-shot optical lithography were obtained at 1 nm intervals from 480 nm to 700 nm using a confocal laser scanning optical microscope (FV3000, Olympus) at an excitation wavelength of 405 nm to generate PL peak wavelength mapping data, of Fig. 1D and Fig. 2B in the main text. Fluorescent images of Fig. 2D, E and Fig. 3C, D were obtained using a fluorescence microscope (BX51, Olympus) with a UV light from an Ag lamp.

## Demonstrations of compressive optical spectroscopy with multivariant emissive films

The composition-graded CsPbBr$_x$Cl$_{3-x}$ and CsPbI$_x$Br$_{3-x}$ alloy films were used to generate PL with the peak wavelength varying over a broad range by linearly moving the films while illuminating a 405 nm confocal laser beam. The excitation power is 36 μW, and the PL integration time is 0.5 s. A grating-based spectrometer and CCD camera used to obtain the PL spectra, yielding a spectral resolution of approximately 0.53 nm near $\lambda_{PL}$ of 450 nm. Repeating the scan confirmed that identical series of PL spectra were obtained, demonstrating the stability of the perovskite films under measurement conditions. The series of PL spectra is presented in a *m*-by-*n* transfer matrix, *A*, with rows and columns corresponding to the excitation position and $\lambda_{PL}$, respectively. To demonstrate a dispersive optics-free spectrometer, excitation position-dependent $I_{PL}$ were measured by a single-cell photodetector after a sample of interest without resolving the spectral information. In principle, the measured $I_{sample}(a)$ is equal to $A \cdot T_{sample}(\lambda)$, where $T_{sample}(\lambda)$ represents an *n*-by-1 matrix for the optical transmittance spectra of the sample. To deduce $T_{sample}(\lambda)$, we conducted a numerical minimization process for $A \cdot T_{sample}(\lambda) - I_{sample}(a)$ using the Nelder–Mead method.

**Estimation of photocatalytic efficiency for halide exchange, $\alpha$.**

The number of exchanged halogens per photo-excited carriers in the host film can be expressed as follows.

$$d_h \frac{dX_H}{dN_{ex}} = \alpha(1 - X_H)$$

, where $d_h$ is the areal density of the halogen elements at the surface, $X_H$ is the relative ratio of substituted halogen among all the halogens, $N_{ex}$ is the photo-excited carrier density at the surface of the host film and $\alpha$ is a coefficient. For simplicity, we assume that all photo-excited carriers are located on the surface, and there are enough precursors in the environment, not limiting the total exchange rate. The surface of perovskite is assumed to be (100) crystalline planes. Then, $X_H$ can be expressed as a function of $N_{ex}$ as follows.

$$X_H = 1 - e^{-\frac{\alpha N_{ex}}{d_h}}$$

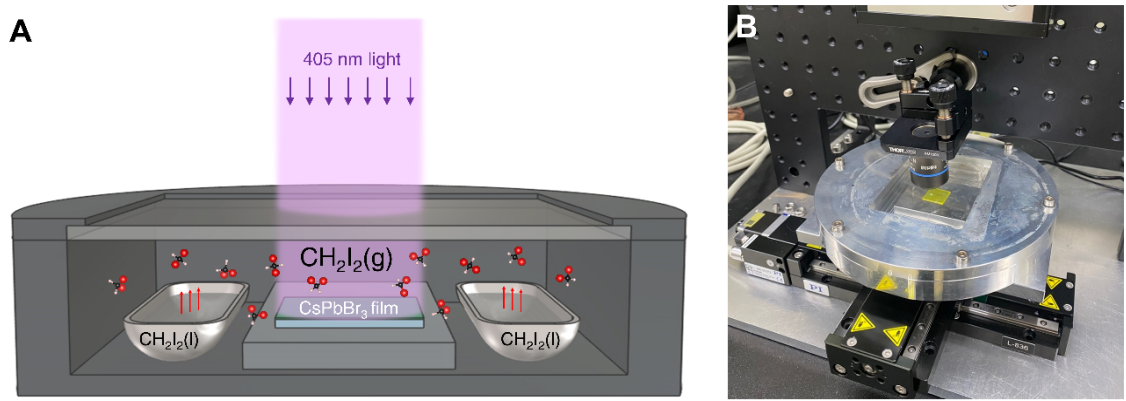

**Figure S1.** (A) Schematics of the system, where the gas environment is controlled by the loaded precursors. Semiconductor alloy films are loaded inside the optical chamber, where UV lights are illuminated through a density-graded optical mask. The illumination area can be modulated from the diffraction limit in a confocal set-up to global illumination with a beam expander. (B) Photograph of the system for manufacturing multispectral films. The UV light is focused by an optical lens and the chamber is moved by a moving stage to change its position.

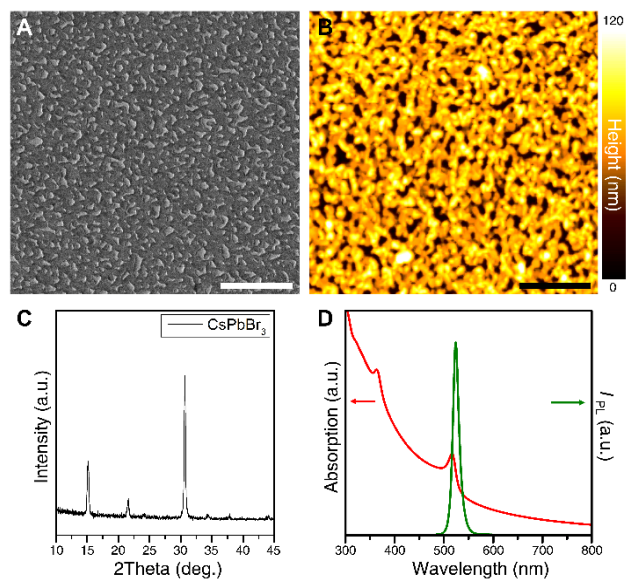

**Figure S2.** (A) SEM image. Scale bar: 20 μm. (B) AFM height map, showing an average height of ~100 nm. Scale bar: 10 μm. (C) XRD spectrum, presenting the cubic structure of CsPbBr$_3$[1]. (D) Optical absorption (red) and PL (green) spectra.

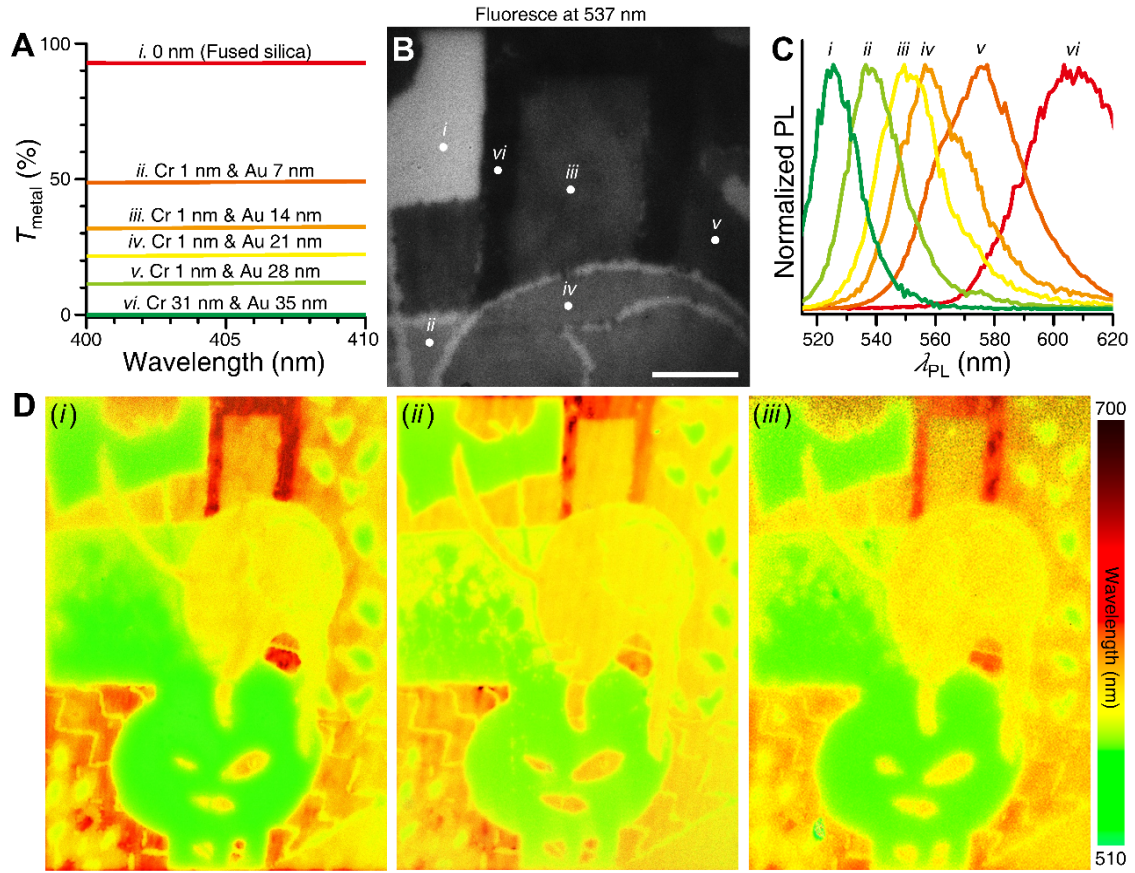

**Figure S3.** (A) Optical transmission spectra through different positions of the photomask with varying thicknesses of metal. (B) Fluorescence image with a bandpass filter (center wavelength: 537 nm, bandwidth: 1 nm). Scale bar: 200 μm. (C) PL spectra from the positions marked in (B). (D) PL peak position maps of reproduced $CsPbI_xBr_{3-x}$ films with the same pattern by repeating the same lithography process.

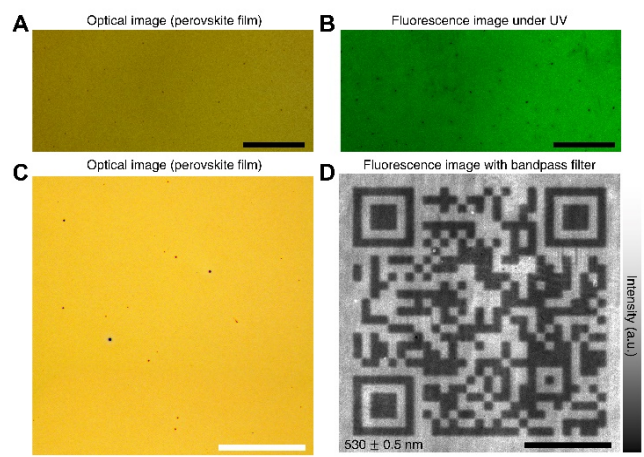

**Figure S4.** (A) Optical image and (B) fluorescence image measured with a Bayer filter of the patterned CsPbCl$_x$Br$_{3-x}$ perovskite film shown in Fig. 3A-D. Scale bar, 50 µm. (C) Optical image with a Bayer filter and (D) fluorescence image with a bandpass filter (center wavelength is 530 nm and band width is 1nm) for QR code patterned CsPbI$_x$Br$_{3-x}$ perovskite film shown in Fig. 4. Scale bar, 500 µm